\documentclass[preprintnumbers,amsmath,amssymb,floatfix,11pt,prd,onecolumn,showpacs,superscriptaddress,nofootinbib]{revtex4}
\usepackage{graphicx}
\usepackage{epsfig}
\usepackage{bm}
\usepackage{amsfonts}

\begin{document}

\title{ Energy conditions in generalized teleparallel gravity models}

\author{\textbf{ Mubasher Jamil}} \email{jamil.camp@gmail.com}
\affiliation{Center for Advanced Mathematics and Physics (CAMP),\\
National University of Sciences and Technology (NUST), H-12,
Islamabad, Pakistan}
\affiliation{Eurasian International Center
for Theoretical Physics,  L.N. Gumilyov Eurasian National University, Astana
010008, Kazakhstan}
\author{\textbf{ D. Momeni}}
\email{d.momeni@yahoo.com }
 \affiliation{Eurasian International Center
for Theoretical Physics,  L.N. Gumilyov Eurasian National University, Astana
010008, Kazakhstan}
\author{\textbf{ Ratbay Myrzakulov}}
\email{rmyrzakulov@gmail.com}\affiliation{Eurasian International Center
for Theoretical Physics,  L.N. Gumilyov Eurasian National University, Astana
010008, Kazakhstan}

\begin{abstract}
{\bf Abstract:}
 In this paper, we investigate the energy conditions (including null, weak, strong, dominant) in generalized teleparallel gravities including pure $F(T)$, teleparallel gravity with  non-minimally coupled scalar field and $F(T)$  with  non-minimally coupled scalar field models. In particular, we apply them to  Friedmann-Robertson-Walker (FRW) cosmology and obtain some corresponding results. Using two specific phenomenological forms of $F(T)$, we show that some of the energy conditions are violated.

\end{abstract}

 \maketitle

\newpage
\section{Introduction}


There are four candidates of the gravitational sector of model: (A)
metric compatible and torsionless (Einstein gravity); (B) Weyl's
type and torsionless (Weyle space); (C) metric compatible with
torsion (Einstein-Cartan space); and (D) Weyl's type with torsion
(Einstein-Cartan-Weyl theory).  A recent version of torsion based
gravity is $F(T)$ (commonly termed `generalized teleparallel
gravity') which is based on the Einstein-Cartan geometry
\cite{f(T)}, where $T$ is the torsion scalar constructed from the
tetrad.  Choosing $F(T)=T$, leads to the pure teleparallel gravity
\cite{hayashi,hehl} which is in good agreement with some standard
tests of the general relativity at the solar system scale
\cite{hayashi}. Numerous features of theoretical interest have been
studied in this gravity already including Birkhoff's theorem
\cite{birkhoff}, cosmological perturbations \cite{zheng},
cosmological attractor solutions \cite{momeni}, generalized second
law in $F(T)$ \cite{gsl} and phantom crossing of the state parameter
\cite{bamba}. Moreover, the local Lorentz invariance is violated
which henceforth leads to violation of first law of thermodynamics
\cite{miaoli,sotirio}. Also the entropy-area relation in this
gravity takes a modified form \cite{gsl}. The Hamiltonian structure
of $F(T)$ gravity has been investigated  and found that there are
five degrees of freedom \cite{miao}.


In teleparallel gravity, the equations of motion for any geometry
are exactly the same as of general relativity. Due to this reason,
the teleparallel gravity is termed as `teleparallel equivalent of
general relativity' \cite{sharif}. In teleparallel gravity, the dark
energy puzzle is studied by introducing a scalar field with a
potential. If this field is minimally coupled with torsion, then
this effectively describes quintessence dark energy. However if it
is non-minimally coupled with torsion, than more rich dynamics of
the field appears in the form of either quintessence or phantom
like, or by experiencing a phantom crossing \cite{geng}. Xu et al
\cite{xu} investigated the dynamics and stability of a canonical
scalar field non-minimally coupled to gravity (arising from
torsion). They found that the dynamical system has an attractor
solution and rich dynamical behavior was found. In the context of
general relativity, a scalar field non-minimally coupled with
gravity has been studied in \cite{mark}.

In this paper, we discuss the energy conditions  in generalized
teleparallel gravities including pure $F(T)$, teleparallel gravity
with  non-minimally coupled scalar field and $F(T)$  with
non-minimally coupled scalar field models. In particular, we apply
them to  FRW cosmology and obtain some corresponding results. Using
two specific phenomenological forms of $F(T)$, we show that some of
the energy conditions are violated.


We follow the plan: In section II we propose three  phenomenological
models for DE with scalar fields and torsion. In section III, we
provide conclusion.

\section{Energy conditions and Generalized teleparallel gravities}

Using the modified (effective) gravitational field equations the null energy condition
(NEC), weak energy condition (WEC), strong energy condition (SEC) and the dominant energy
condition (DEC) are given by \cite{lobo,anzhong}
\begin{eqnarray}
\text{NEC}&\Longleftrightarrow&\rho_{\text{eff}}+p_{\text{eff}}\geq0.\label{n1}\\
\text{WEC}&\Longleftrightarrow& \rho_{\text{eff}}\geq0\ \text{and}\ \rho_{\text{eff}}+p_{\text{eff}}\geq0.\label{n2}\\
\text{SEC}&\Longleftrightarrow& \rho_{\text{eff}}+3p_{\text{eff}}\geq0\ \text{and}\ \rho_{\text{eff}}+p_{\text{eff}}\geq0.\label{n3}\\
\text{DEC}&\Longleftrightarrow& \rho_{\text{eff}}\geq0\ \text{and}\ \rho_{\text{eff}}\pm p_{\text{eff}}\geq0.\label{n4}
\end{eqnarray}
The origin of these energy conditions is independent of any gravity theory and that these are purely geometrical \cite{hawking}. Note that NEC implies WEC and WEC implies SEC and DEC. In all the subsequent models we will assume that the regular matter  satisfies all the energy conditions separately i.e. $\rho_m\geq0$, $\rho_m\pm p_m\geq0$, $\rho_m+3p_m\geq0$. In literature, the $f(R)$ theory has been tested against the energy conditions \cite{wang1}. Thus, we need to check the validity of these conditions for energy density and pressure for the torsion and scalar field.

\subsection{$T+f(T)$ model }

The action of a gravity model based on pure but arbitrary torsion with a matter field is given by  \cite{wei}
\begin{equation}\label{S'}
\mathcal{S}=\int d^4x~e(T+f(T)+\mathcal{L}_m),
\end{equation}
where $e=\text{det}(e^i_\mu)=\sqrt{-g}$, where $e_i(x^\mu)$ are related to the metric via $g_{\mu\nu}=\eta_{ij}e_\mu^i e^j_\nu$, where all indices run over 0,1,2,3.
The Friedmann equations in effective notation are given by
\begin{eqnarray}
\rho_\text{eff}=3H^2, \ \
p_\text{eff}=-(2\dot H+3H^2),
\end{eqnarray}
where
\begin{eqnarray}
\rho_\text{eff}&=&\rho_m+\rho_T\nonumber\\&&=\rho_m+Tf_T-\frac{f}{2},\label{e1}\\
p_\text{eff}&=& p_m+p_T\nonumber\\&&=p_m+(2\dot H-T)f_T+4\dot HTf_{TT}+\frac{f}{2}.\label{e2}
\end{eqnarray}
To check the viability of this cosmological model, we check the energy conditions (\ref{n1})-(\ref{n4}) using (\ref{e1}) and (\ref{e2}):
\begin{eqnarray}
2\dot H( f_T +2Tf_{TT})\geq0\\
 -f+2Tf_T\geq0\\
 f+2Tf_T+6\dot H(f_T+2Tf_{TT})\geq0\\
2\dot H(f_T+2Tf_{TT})+(f-2Tf_T)\leq0
\end{eqnarray}

Now we use two recently proposed models of $f(T)$ gravity \cite{puxun}
\begin{eqnarray}
f_1(T)&=&\alpha (-T)^n \tanh\Big(\frac{T_0}{T}\Big)\nonumber\label{f1}\ \ \ \ \ \ \text{(Model-I)} \\
f_2(T)&=&\alpha (-T)^n \Big[1-\exp\Big(-p\frac{T_0}{T}\Big)\Big]\label{f2}\nonumber\ \ \ \ \text{(Model-II)}
\end{eqnarray}
Here $T_0=-6H_0^2$. These two models are able to give rise to crossing the phantom divide. In Model-I the exponent $n>\frac{3}{2}$, the parameter of the model \cite{puxun}
$$
\alpha=-\frac{1-\Omega_{m0}-\Omega_{r0}}{(6H_0^2)^{n-1}\Big[\frac{2}{\cosh(1)^2}+(1-2n)\tanh(1)\Big]}
$$
Similarly for Model-II, we know that $n>\frac{1}{2}$ and \cite{puxun}
$$
\alpha=\frac{(6H_0^2)^{1-n}(1-\Omega_{m0}-\Omega_{r0})}{-1+2n+e^p(1-2n+2p)}
$$
Here $\Omega_{m0},\Omega_{r0}$ are the present values of the energy densities of the dark matter and radiation.
Now we examine the energy conditions, based on these two viable models.

\begin{itemize}

\item NEC: First note that the NEC reduces to
\begin{equation}
-4\dot H\sqrt{-T}\frac{d}{dT}\Big[\sqrt{-T}f_T\Big]\leq0
\end{equation}
For model-I in case $\dot H>0$, the top left figure shows this model does not satisfy the NEC but in case $\dot H<0$ this model satisfies NEC. For model-II similarly, when $\dot H>0$ the NEC breaks down but when $\dot H<0$, NEC is valid (top right figure).

\item WEC: It is easy to show that we must check the following inequality

\begin{equation}
\frac{d}{dT}\Big[\frac{f(T)}{\sqrt{-T}}\Big]\leq 0
\end{equation}
 For model-I , middle left figure shows that the WEC is satisfied. Also for model-II, this  energy condition will be satisfied as can be seen in middle right figure.

\item SEC: In regime $\dot H>0$, we know from NEC $f_T+2Tf_{TT}\geq 0$.  It remains only to check whether $f+2Tf_T\geq 0$. Note that  $f+2Tf_T=-2\sqrt{-T}\frac{d}{dT}\Big(f\sqrt{-T}\Big)\geq0$.
For models I and II, we observe that this SEC is satisfied as shown in bottom left and right figures.

\item DEC: We can write this condition as $$-4\dot H \sqrt{-T}\frac{d}{dT}(\sqrt{-T}f_T)-2(-\frac{f}{2}+Tf_T)\leq0 .$$ Note that the last bracket is positive on account of validity of WEC. Also from NEC, we know that for both models $f_1$ and $f_2$, always $\frac{d}{dT}(\sqrt{-T}f_T)\geq0$. If $\dot H>0$ then DEC is satisfied.

\end{itemize}

\begin{figure*}[thbp]
\begin{tabular}{rl}
\includegraphics[width=7.5cm]{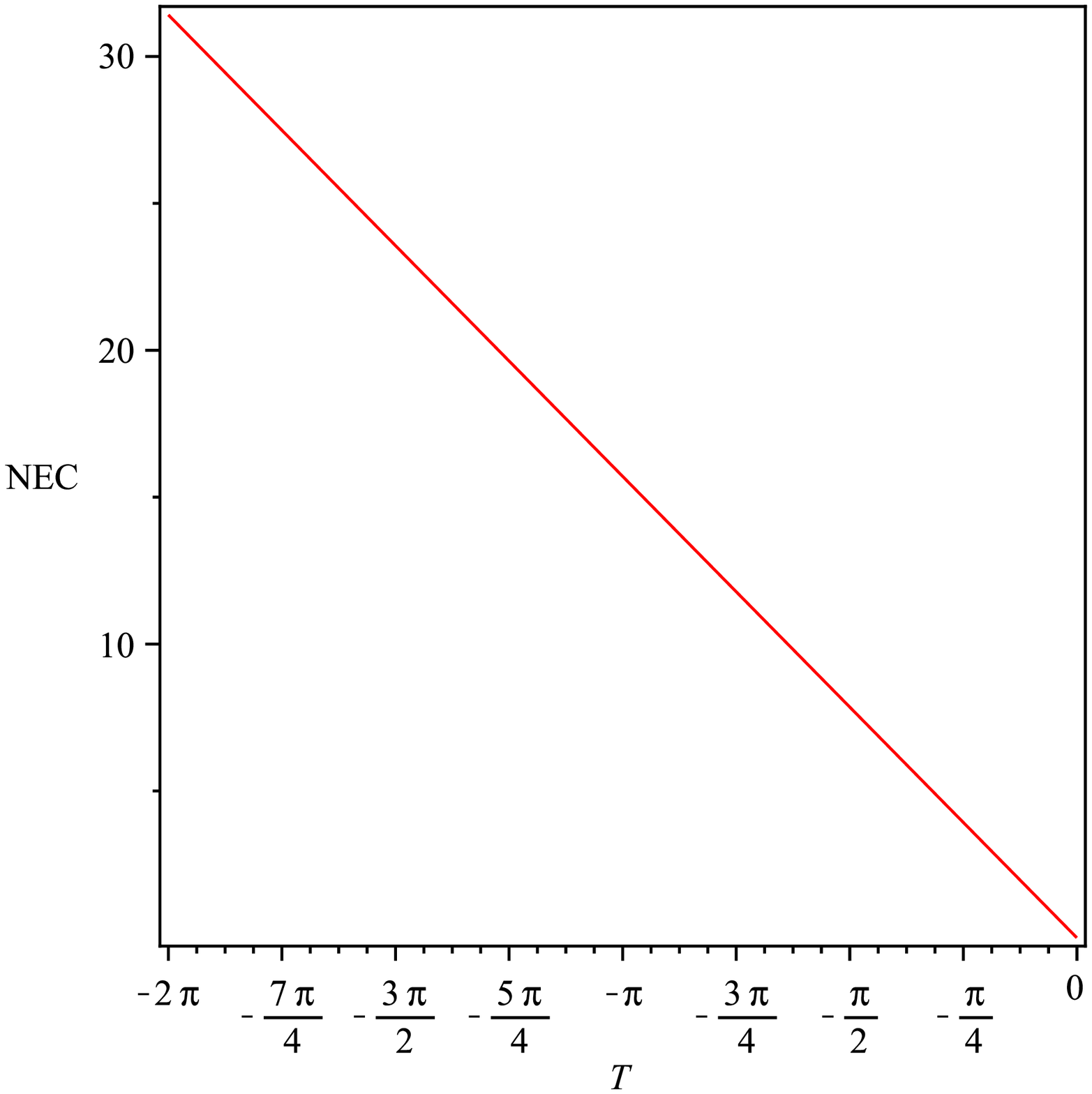}&
\includegraphics[width=7.5cm]{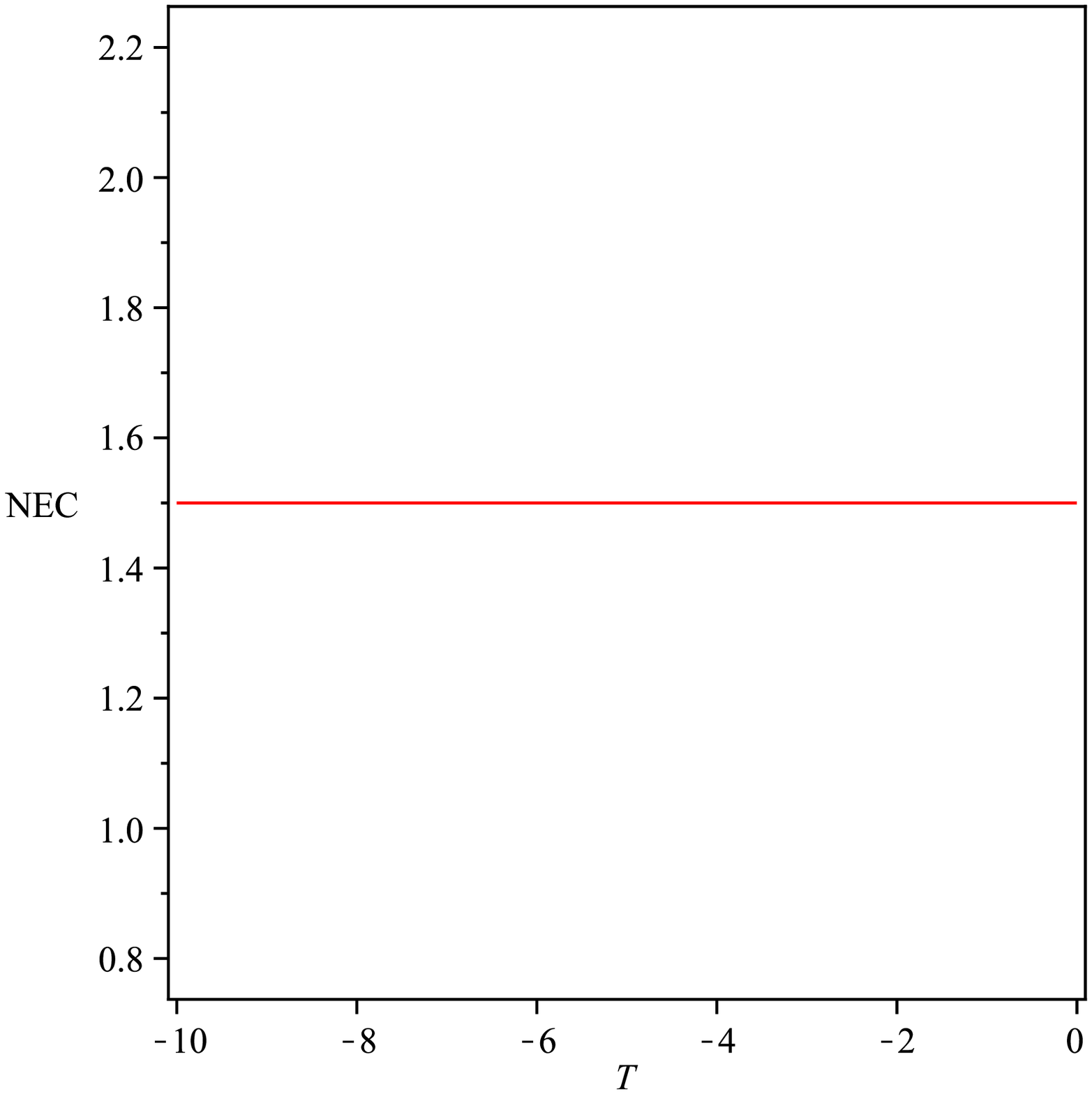} \\
\includegraphics[width=7cm]{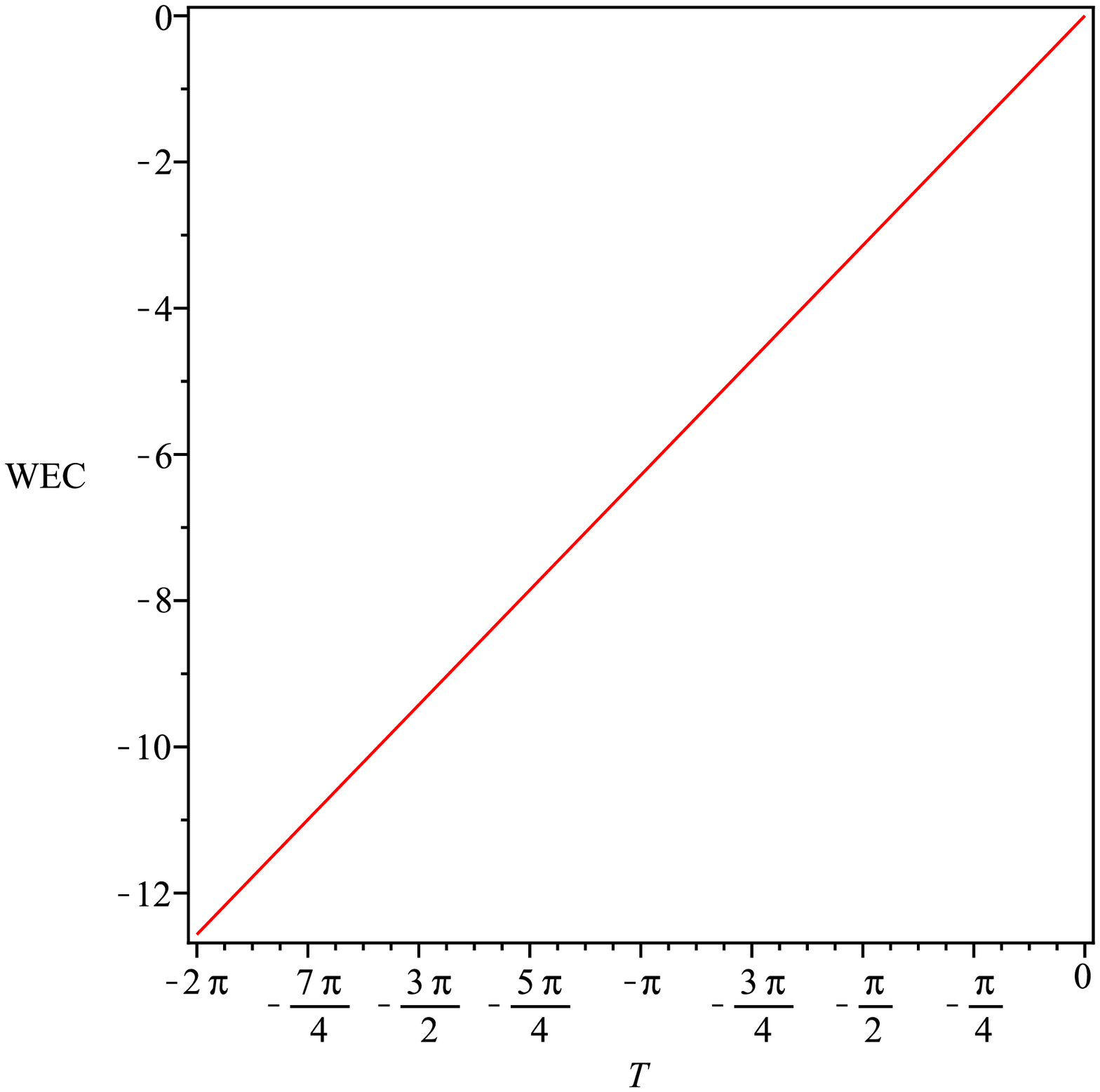}&
\includegraphics[width=7cm]{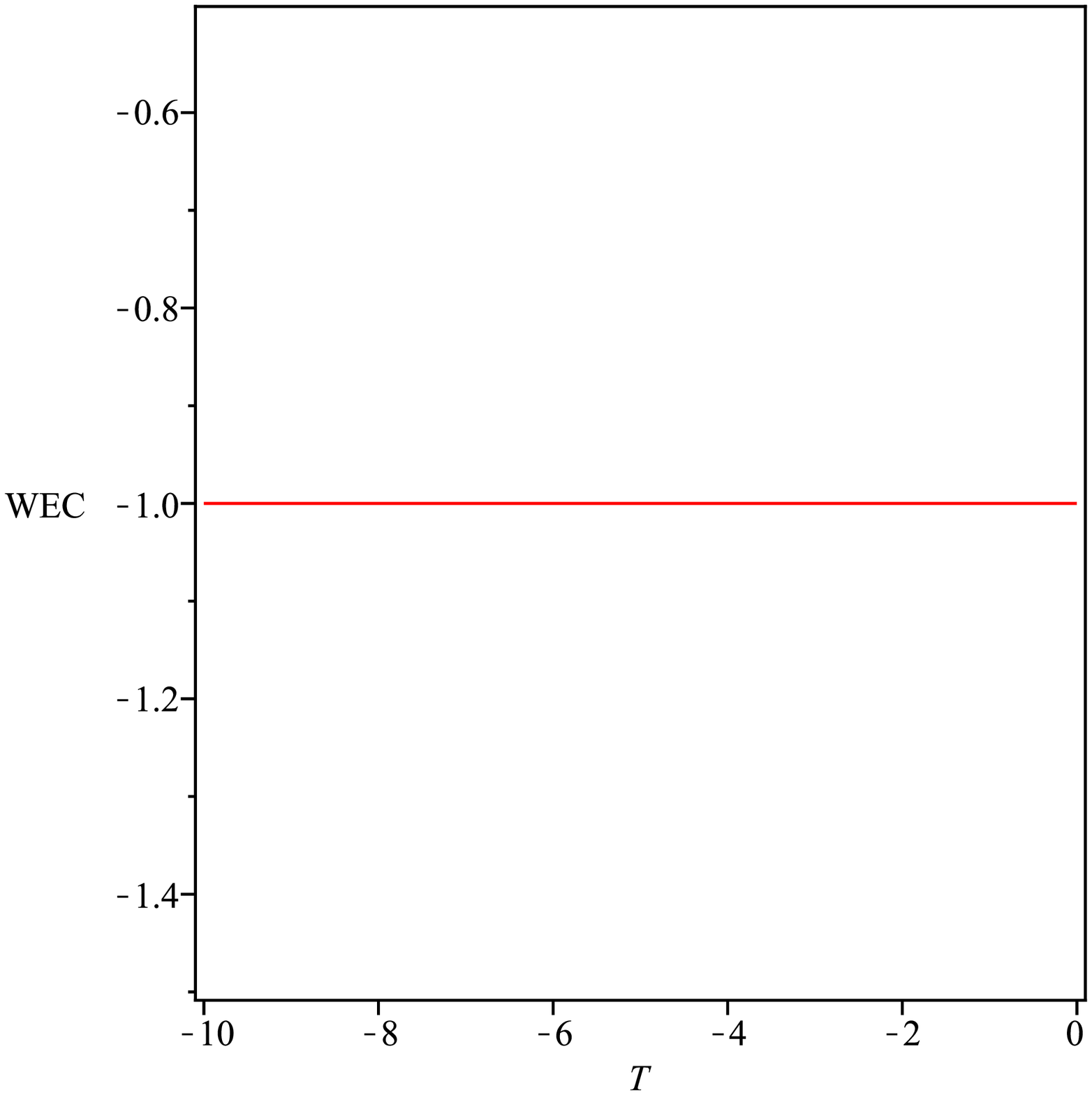} \\
\includegraphics[width=7cm]{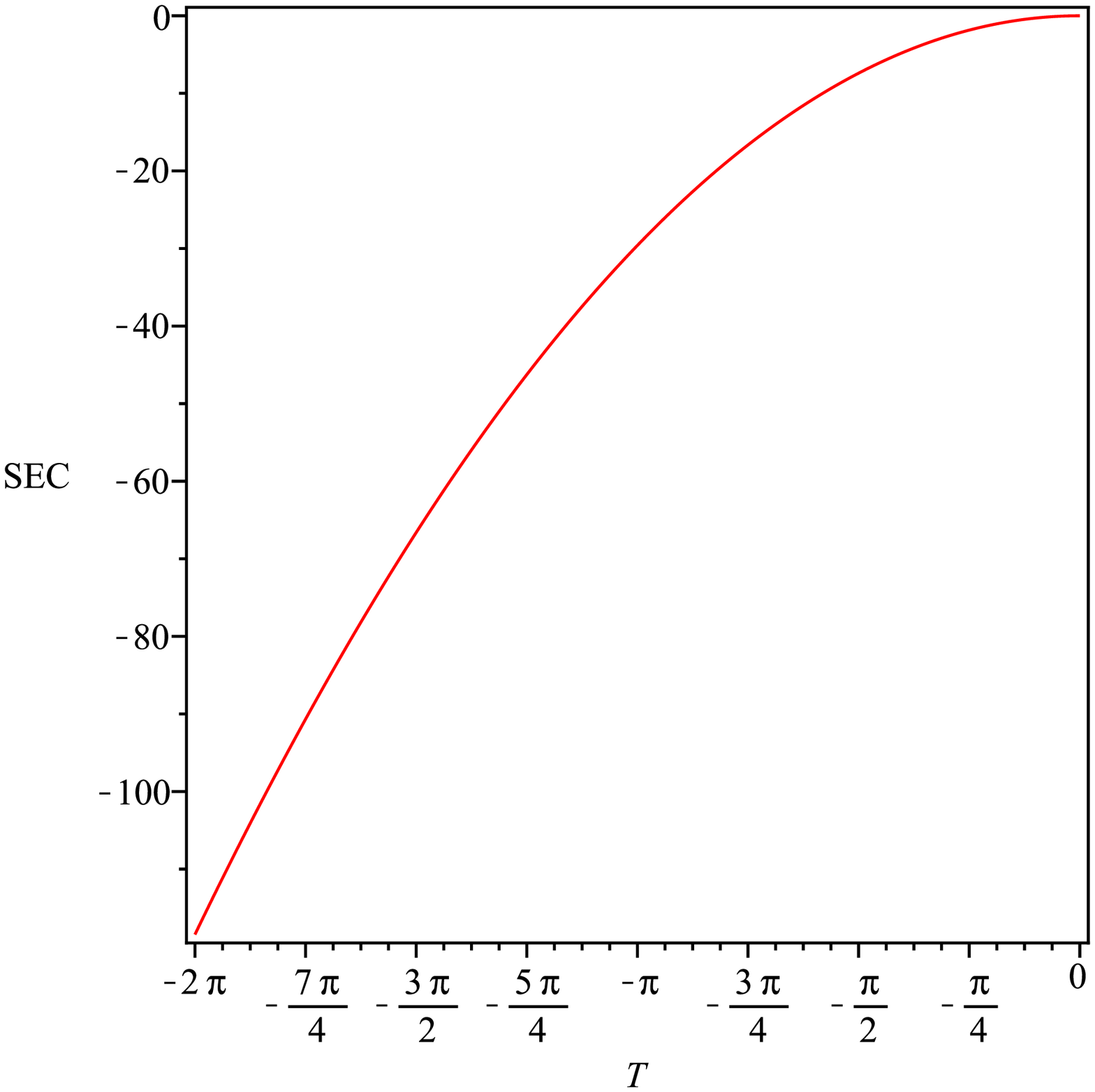}&
\includegraphics[width=7cm]{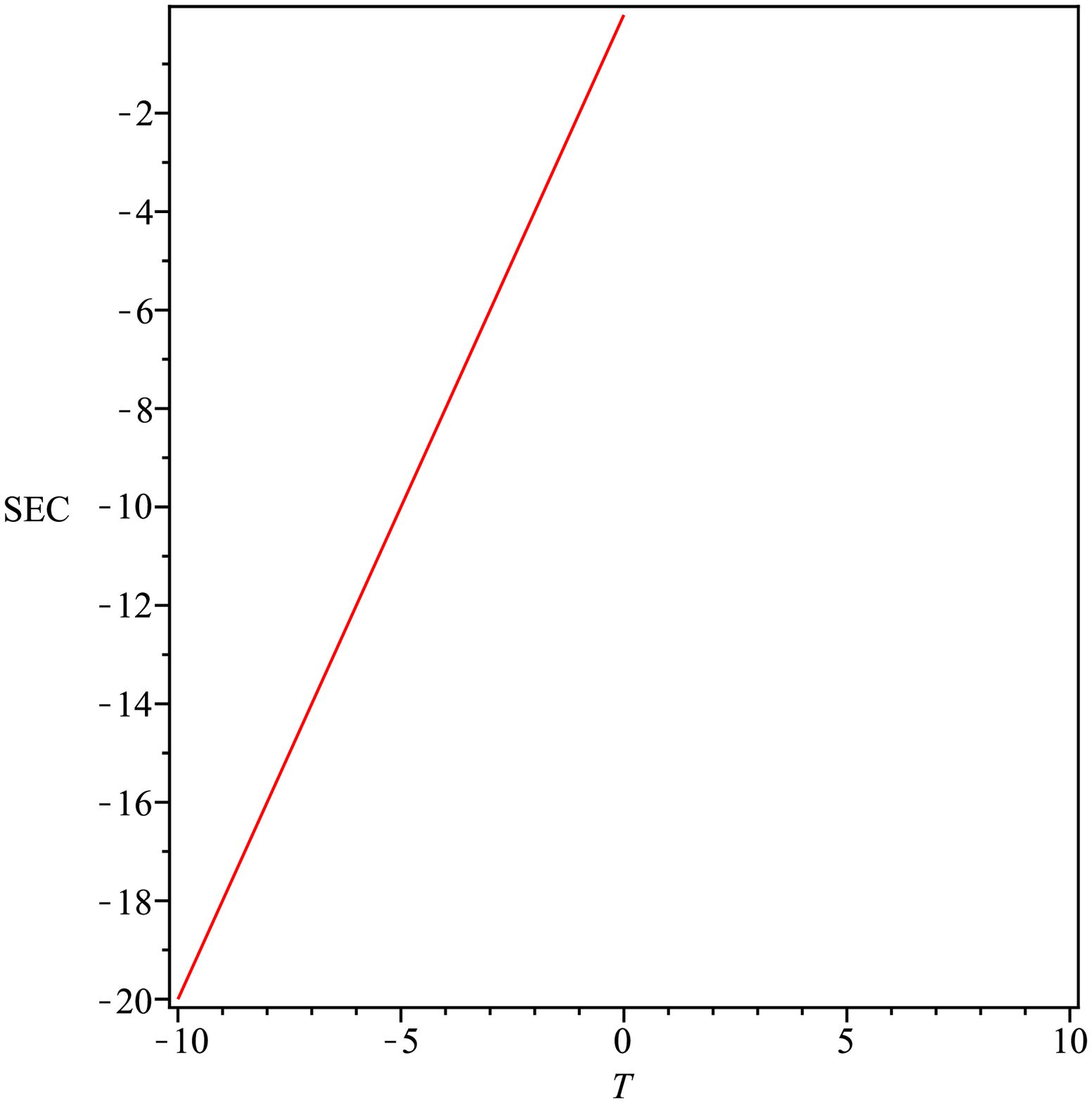} \\
\end{tabular}
\caption{ (\textit{Top Left})  NEC for model 1.   (\textit{Top Right}) NEC for model 2. (\textit{Middle Left}) WEC for model 1. (\textit{Middle Right}) WEC for model 2.  (\textit{Bottom Left}) SEC for model 1. (\textit{Bottom Right})SEC for model 2. The left and right panel figures correspond to $f_1$ and $f_2$, respectively.  }
\end{figure*}

\subsection{Teleparallel Gravity with a non-minimally coupled scalar field}

As a generalization of pure teleparallel gravity, Xu and collaborators \cite{xu} added a canonical scalar field,
allowing for a nonminimal coupling with (teleparallel) gravity. Their proposed action with a little modification is
\begin{eqnarray}
\mathcal{S}&=&\int  d^4x a^3\Big[ \frac{T}{2}+\frac{1}{2}(\epsilon\phi_{,\mu}\phi^{,\mu}+\xi T\phi^2)\nonumber\\&&-V(\phi)+\mathcal{L}_m  \Big].
\end{eqnarray}
The symbol $\epsilon=+1,-1$ refer to quintessence and phantom dark energy respectively. $\xi$ represents the coupling between the scalar torsion and the scalar field. $\mathcal{L}_m$ is the Lagrangian density for the matter component. In this gravity, the effective energy density and effective pressure for a flat FRW universe is \cite{xu}
\begin{eqnarray}
\rho_{\text{eff}}&=&\rho_m+\frac{\epsilon}{2}\dot\phi^2+V(\phi)-3\xi H^2\phi^2,\\
p_{\text{eff}}&=&p_m+\frac{\epsilon}{2}\dot\phi^2-V(\phi)+4\xi H\phi\dot\phi\nonumber\\&&+\xi\phi^2(3H^2+2\dot H).
\end{eqnarray}

\begin{itemize}

\item NEC: $\epsilon\dot\phi^2+4\xi H\phi\dot\phi+2\xi\dot H\phi^2\geq0$. We discuss two cases:\\
Case-1: $\epsilon=+1,\ \ \dot H<0$: In this case NEC is obeyed when $\dot H \geq-\frac{1}{2\xi \phi^2}\Big(\dot\phi^2+4\xi \phi \dot \phi H\Big)$.\\
Case-2: $\epsilon=-1,\ \ \dot H>0$: In this case NEC is justified when $\dot \phi^2\leq 4\xi H \phi \dot\phi+2\xi\dot H \phi^2$.

\item WEC: If $\epsilon=+1$, $V(\phi)\geq0$, $3\xi H^2 \phi^2\leq \frac{\epsilon}{2}\dot\phi^2+V(\phi)$.
If $\epsilon=-1$, $V(\phi)\geq0$, $ V(\phi)\geq 3\xi H^2 \phi^2+ \frac{\epsilon}{2}\dot\phi^2$.

\item SEC:  We must check whether $\epsilon \dot\phi^2-V(\phi)+3\xi H^2\phi^2+6\xi H\phi \dot\phi+3\xi \dot H \phi^2\geq0$.\\
Case-1: $\epsilon=+1,\ \ \dot H<0$: In this quintessence model, we must have
$\dot\phi\geq\frac{1}{6\xi H \phi}\Big(-\dot\phi^2+V(\phi)-3\xi H^2\phi^2-3\xi \dot H \phi^2\Big)$.\\
Case-2: $\epsilon=-1,\ \ \dot H>0$: We conclude that $\dot \phi\geq\frac{1}{6\xi\phi H}\Big(\dot\phi^2+V(\phi)-3\xi H^2\phi^2-3\xi \dot H \phi^2\Big)$.

\item DEC: First we check the condition $\rho_{\text{eff}}\geq0$. It means that $\frac{\epsilon}{2}\dot\phi^2+V(\phi)-3\xi H^2\phi^2\geq0$. We have two special cases\\
Case-1: $\epsilon=+1$. In this case $\dot\phi^2\geq2(3\xi H^2\phi^2-V(\phi))$.\\
Case-2: $\epsilon=-1$. Here $\dot\phi^2\leq2(3\xi H^2\phi^2-V(\phi))$.\\
Further the condition $\rho_{\text{eff}}\geq p_\text{eff}$ we have\\
$$
\dot\phi\leq\frac{2V(\phi)-6\xi H^2\phi^2-2\xi\phi^2\dot H}{4\xi H \phi}
$$

\end{itemize}

\subsection{$T+f(T)$ model with a minimally coupled scalar field}

 The total action with contributions from torsion, matter and a minimal coupled scalar field component reads
\begin{equation}\label{S}
\mathcal{S}=\int d^4xa^3\Big[F(T)-\rho_m+\frac{1}{2}\epsilon \phi_{,\mu}\phi^{,\mu}-V(\phi)\Big].
\end{equation}
where $F(T)=T+f(T)$. The scalar field $\phi$ has the potential energy $V(\phi)$  and $\rho_m=\rho_{m0}a^{-3}$ is the energy density of matter with vanishing pressure and $\rho_{m0}$ is a constant energy density at some initial time.

 The forms of effective energy density and pressure are
\begin{eqnarray}
\rho_\text{eff}&=&\rho_m+\rho_\phi+\rho_T\nonumber\\&&=\rho_m+\frac{1}{2}\epsilon\dot\phi^2+V(\phi)+Tf_T-\frac{f}{2},\\
p_\text{eff}&=& p_m+p_\phi+p_T\nonumber\\&&=p_m+\frac{1}{2}\epsilon\dot\phi^2-V(\phi)+(2\dot H-T)f_T\nonumber\\&&+4\dot HTf_{TT}+\frac{f}{2}.
\end{eqnarray}
The analysis of energy conditions for this model is given below:
\begin{itemize}

\item NEC: $\epsilon\dot\phi^2+2\dot H( f_T +2Tf_{TT})\geq0$. Notice that the last term in bracket is positive as demonstrated in section A. If $\epsilon=+1$,  then NEC holds but violated otherwise.

\item WEC: If $\epsilon=+1$ and $V(\phi)>0$, than from section A, we have $-\frac{f}{2}+Tf_T\geq0$. Thus WEC holds. If $\epsilon=-1$ and $V(\phi)>0$, then WEC is satisfied if $V\geq\dot\phi^2+\frac{f}{2}-Tf_T. $

\item SEC: If $\epsilon=+1$, $\dot H<0$ and $V(\phi)>0$, than $2\dot\phi^2+f\geq 2V(\phi)-2Tf_T-6\dot H f_T$. If $\epsilon=-1$, $\dot H>0$ and $V(\phi)>0$, than $f+6\dot H f_T\geq 2\dot\phi^2+ 2V(\phi)-2Tf_T$

\item DEC: If $\epsilon=+1$, $\dot H<0$ and $V(\phi)>0$, than $\frac{1}{2}\dot\phi^2+V(\phi)\geq \frac{f}{2}-Tf_T$. If $\epsilon=-1$, $\dot H>0$ and $V(\phi)>0$, than $V(\phi)\geq \frac{f}{2}-Tf_T+\frac{1}{2}\dot\phi^2.$

\end{itemize}

\section{Conclusion}

In this paper, we discussed the energy conditions in  three
different models of generalized teleparallel gravities. These energy
conditions are stronger test to check the viability of these
theories. Here we examined three torsion based models with two
phenomenological forms of $f(T)$. In the case of pure $f(T)$
gravity, we showed that all the energy conditions can be satisfied
for both kind of dark energy models. Then by adding a non-minimally
scalar interaction, we showed that these energy conditions can be
fulfilled for some specific values of the dynamical quantities of
the model. Further we showed that given a minimally coupled dark
energy component depending on the value of interaction, the energy
conditions can be valid.



\end{document}